\documentclass[10pt,journal,compsoc]{IEEEtran}
\usepackage[utf8]{inputenc}
\usepackage{cite}
\usepackage{url}
\usepackage{multirow}
\usepackage{threeparttable}
\usepackage{graphicx}
\usepackage{booktabs}
\usepackage{subfigure}
\usepackage{amssymb}
\usepackage{pifont}
\usepackage{tikz}
\usepackage{enumitem}

\newcommand{\cmark}{\ding{52}}%
\newcommand{\xmark}{\ding{56}}%
\newcommand*\circled[1]{\tikz[baseline=(char.base)]{
    \node[shape=circle,draw,inner sep=0.5pt] (char) {#1};}}

\title{A Quantitative Survey of Communication Optimizations in Distributed Deep Learning}
	
\author{Shaohuai Shi, 
    Zhenheng Tang,
    Xiaowen Chu\IEEEauthorrefmark{1}\thanks{*Corresponding author.},
    Chengjian Liu,
    Wei Wang,
    and Bo Li
    \IEEEcompsocitemizethanks{
	\IEEEcompsocthanksitem Shaohuai Shi, Wei Wang and Bo Li are with the Department of Computer Science and Engineering, The Hong Kong University of Science and Technology.
	\IEEEcompsocthanksitem Zhenheng Tang and Xiaowen Chu are with the Department of Computer Science, Hong Kong Baptist University. 
	\IEEEcompsocthanksitem Chengjian Liu is with the College of Big Data and Internet, Shenzhen Technology University.
	}
}

\begin{document}

\maketitle
\begin{abstract}
 Nowadays, large and complex deep learning (DL) models are increasingly trained in a distributed manner across multiple worker machines, in which extensive communications between workers pose serious scaling problems. In this article, we present a quantitative survey of communication optimization techniques for data
 parallel distributed DL. We first identify the major communication challenges and classify the existing solutions into three levels, namely the learning algorithm, the system architecture, and the network infrastructure. We present the state-of-the-art communication optimization techniques and conduct a comparative study of seven common lossless distributed DL methods on a 32-GPU cluster with 100Gbps InfiniBand (IB). We show that (1) the DL models with low model intensity (such as BERT and BERT-Large) are difficult to scale out even with the best available lossless algorithm over 100Gbps IB; (2) the system architecture and scheduling algorithms have a critical impact on the scaling property. We conclude the article with discussions on the open issues for further investigations.
 
\end{abstract}

\section{Introduction}
\label{sec:intro}

The remarkable technological advances of deep learning (DL) have enabled a
multitude of practical AI applications, ranging from computer vision to
natural language processing and to robotics. In a typical DL workflow, deep
neural network models are trained to solve a learning problem
(e.g., image classification) on a labeled dataset; the trained models can
then be used to make an inference given a new input (e.g., predicting the
image label). Popular DL training algorithms include the standard mini-batch
stochastic gradient descent (SGD) and its variants. These algorithms minimize
a pre-defined loss function by iteratively updating the model parameters with
stochastic gradients, calculated by sampling a mini-batch of data from the
training set. 

According to a recent study from OpenAI, the computational complexity required
in DL training has doubled every 3.4 months since 2012, outpacing the Moore's
Law. As the training data and the DL models grow exponentially larger
(e.g., the BDD100K auto-driving dataset has 120 million images, and the
BERT-xlarge language model has over 1 billion parameters), training deep
models on a single GPU or TPU device results in an exceedingly long time. A common practice is to parallelize DL training
across multiple processors\footnote{Throughout this article, worker and
processor are used interchangeably.} that collaboratively update the model parameters. 
However, such distributed training requires iterative communications between processors,
creating a severe performance bottleneck as the improvement of device
interconnections lags far behind the rapidly increased computing power of AI
processors. The result is the limited system scalability, as suggested by the
Amdahl's law. Therefore, how to address the communication bottlenecks in
distributed DL has attracted great attention from both academia and industry
in recent years.

Model parallelism and data parallelism are the two major parallelization
schemes~\cite{dean2012large} that enable multiple processors to
collaboratively train a single model. Model parallelism splits the set of
model parameters and distributes them to all processors, but the high dependency
between different neurons and the unbalanced parameter sizes in deep models
make model parallelism difficult to scale out. Data parallelism, on the other
hand, distributes the computational workload of different data samples to
different processors that share the same set of model parameters. Compared with
model parallelism, data parallelism is more appealing due to its improved
scalability and simpler implementation. In this article, we mainly focus on
data parallelism. 

\begin{figure}[!ht]
	\centering
	\subfigure[Data parallelism]
	{
	\includegraphics[width=0.36\linewidth]{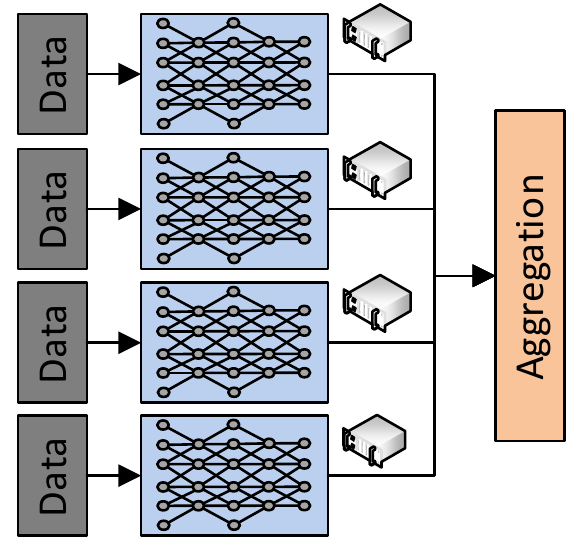}
	}\hspace{5pt}
	\subfigure[A DAG example]
	{
	    \includegraphics[width=0.56\linewidth]{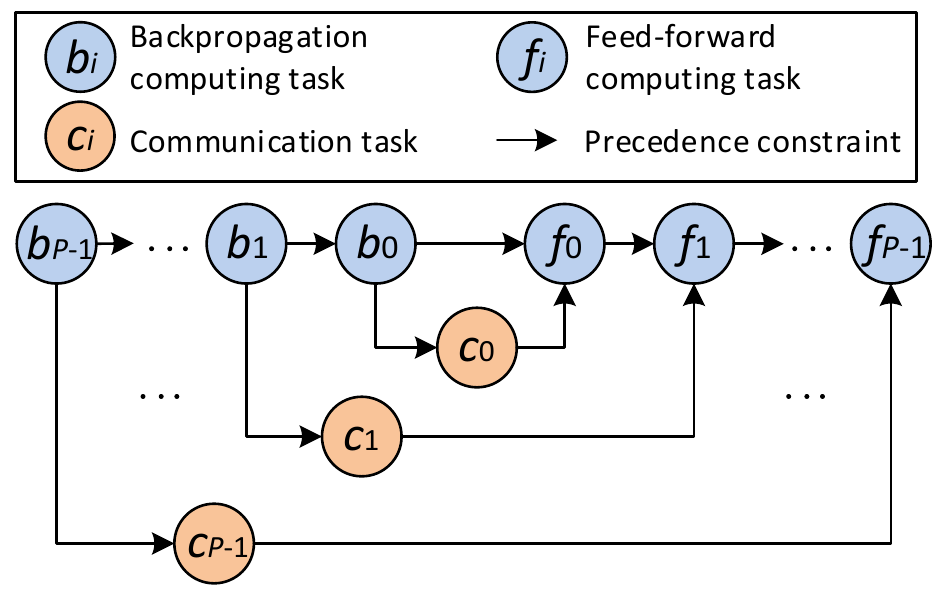}
	}
	\caption{Data parallelism of distributed DL.}
	\label{fig:parallelisms}
\end{figure}

Fig.~\ref{fig:parallelisms}(a) illustrates the popular synchronized
SGD algorithm for distributed DL with data parallelism, which
has the same convergence performance (in terms of the number of iterations) as
SGD on a single worker. In this method, workers load different data samples to
calculate the gradients independently; all gradients are aggregated 
to update the model parameters. Data parallel synchronous SGD can be modeled by a directed acyclic graph (DAG), as shown in
Fig.~\ref{fig:parallelisms}(b). The backpropagation computations of gradients are
from the last layer to the first (denoted by $b_{P-1},...,b_1,b_0$), and the
distributed gradients should be aggregated (denoted by $c_{P-1},...,c_1,c_0$) before going into the feed-forward computations (denoted by $f_0,f_1,
..., f_{P-1}$) of the next iteration. The distributed synchronized SGD is also
known as bulk synchronous parallel (BSP) SGD as it requires communication and
synchronization in every iteration. The gradients can be aggregated through one or more
dedicated parameter servers (PS)~\cite{chen2019round} or by all-to-all (A2A) communications~\cite{shi2019mg}.
 
Much work has been proposed recently to improve the scalability of distributed DL.
In this article, we develop a taxonomy for
describing communication-efficient techniques in distributed DL, and present a
quantitative survey of communication optimization techniques for the BSP-style
training algorithms. We identify the model intensity and batch size as
two key factors that affect the system scalability, and conduct a quantitative
study to compare seven state-of-the-art distributed training methods on a
32-GPU cluster with 100Gbps IB. Our evaluation method and
results can serve as a reference for the practitioners to design their
distributed DL platforms\footnote{Our source code is publicly available at
\url{https://github.com/HKBU-HPML/ddl-benchmarks}.}. Our main observations
through this study are: 
\begin{itemize}
  \item A model with low model intensity and small batch
size (thus a high communication-to-computation ratio) is difficult to scale
out. 
  \item The decentralized A2A architecture is more latency-sensitive than the centralized PS architecture, but the latter requires extra servers and network ports to achieve good performance. 
  \item Scheduling algorithms can be useful to hide the communication costs in both PS and A2A architectures. In particular, tensor fusion is suitable for A2A, while tensor partition is more suitable for PS.
\end{itemize}

The remainder of this article is organized as follows. We first identify the communication issues and existing solutions in distributed DL. Then we elaborate commonly used communication optimization techniques, followed by our experimental study. Finally, we discuss the challenges and possible future research directions.

\section{Communication Issues and Solutions}
\label{sec:issues}
\subsection{Scope, Assumptions, and Terminologies}
In this article, we mainly discuss the communication issues in data parallel distributed DL, and focus on the data center or HPC environments where network speed is high and stable. 

In a typical data parallel distributed DL (e.g., BSP-SGD), each training iteration consists of several steps. First, each worker loads a mini-batch of data as the input and performs feed-forward calculations to calculate the loss value against 
the corresponding labels. Next, each worker backpropagates the loss and
calculates the first-order gradients of model parameters. The local gradients are
aggregated among all workers, and the averaged gradients are
finally used to update the model parameters. The algorithm proceeds to the
next iteration, until a certain convergence condition is met. In this article, we assume data I/O can be overlapped with the computations, and hence will not consider the data I/O time.

Consider a training job of a deep model with $D$ parameters that uses SGD with
a mini-batch size of $M$. Assume the number of arithmetic operations required
for a single data sample in each training iteration is $C$. A data parallelism
solution with $N$ workers will distribute the $MC$ arithmetic operations to
the $N$ workers (e.g., each worker has a \textit{local mini-batch size} of
$M/N$). In the simplest case where communication tasks do not overlap with
computing tasks, the speedup achieved by $N$ workers is
$\frac{t_s}{t_s/N+t_m}$, where $t_s$ is the computing time with a single worker, and $t_m$ is the communication time of distributed training with $N$ workers. As $N$ becomes larger, the
speedup approaches $t_s/t_m$, which explains the
significance of communication optimization in distributed DL. To eliminate the
impact of computing speed and communication speed on the analysis of speedup,
we define \textit{communication-to-computation (C2C) ratio} of a distributed
training job as the total amount of communication traffic divided by the total
amount of computations. Due to the dependency between communication tasks and
computation tasks (Fig. \ref{fig:parallelisms}(b)), C2C ratio is the key
factor that affects the system scalability. 

In practice, the total amount of communication traffic is linearly
proportional to the model size $D$ and also depends on the number of workers
$N$. So we can use $D \cdot f(N)$ to model the amount of
communication\footnote{For simplicity, the unit of communication is the size
of one model parameter or gradient. But in practice, the size of a model
parameter could be different from the size of a gradient.} where $f(N)$
depends on the communication scheme. The C2C ratio can then be calculated by
$\frac{D \cdot f(N)}{M \cdot C}$. We define \textit{model intensity}
$I=\frac{C}{D}$, which is the average number of arithmetic operations in an
iteration per data sample per model parameter. Here, $I$ is an intrinsic feature of
the model that captures the difficulty of parallelism. The C2C ratio can then
be simplified as $\frac{f(N)}{M \cdot I}$. Our experimental results in
Section~\ref{sec:experiments} verify that a model with low intensity $I$
and/or small batch size $M$ is difficult to scale. To
reduce the C2C ratio of a given DL model, we need to design good communication
schemes with small $f(N)$ and choose a large batch size $M$.

\subsection{Communication Issues}
We use the BERT-Large language model (with 336 million parameters) as an example to illustrate the communication challenges in distributed training. Given a local batch size of 8 (which is limited by the available GPU memory size), each iteration requires $597 \times 10^9$ floating point operations (FLOPs) which take 163ms on an Nvidia RTX2080Ti. There are several communication challenges that limit the system scalability of distributed training.

\vspace{.4em}
\noindent \textbf{Communication Size: }In each training iteration, the whole set of model parameters or their gradients should be exchanged across all workers. The BERT-Large model has a size of 1.34GB if the parameters are stored in a 32-bit format. Given $N$ workers, finding the average of $N$ sets of data and synchronizing the updated model within a short time period can be very challenging. For instance, when training BERT-Large on a server with 4 RTX2080Ti connected through PCIe 3.0, each iteration requires 441ms of communication time for the all-reduce operations, resulting in a poor speedup of $1.08\times$.

\vspace{.4em}
\noindent \textbf{Communication Performance: }Deep models have a layered structure, and the parameters and their corresponding gradients are typically stored as tens to hundreds of tensors. First of all, these tensors are calculated layer by layer on the fly, creating intrinsic time dependency that limits the design space of scheduling computing and communication tasks. Second, the tensor size ranges from kilo-bytes to mega-bytes. It is difficult to fully utilize the high network bandwidth when exchanging small messages \cite{shi2019mg}. For example, in our testbed, transmitting 1MB of message across the 10GbE (TCP/IP), 100GbE (TCP/IP), and 100GbIB (RDMA) achieves an effective throughput of 8.2Gbps, 16.5Gbps, and 83.2Gbps, respectively; while transmitting a smaller message of 16KB across the 10GbE, 100GbE, and 100GbIB can only achieve much lower throughput of 1.2Gbps, 4.6Gbps, and 16.7Gbps, respectively. Optimally exchanging various tensors among a set of workers requires a co-design of message exchange algorithm and network system architecture that considers both bandwidth and communication latency.

\begin{figure}[!ht]
	\centering
	\includegraphics[width=\linewidth]{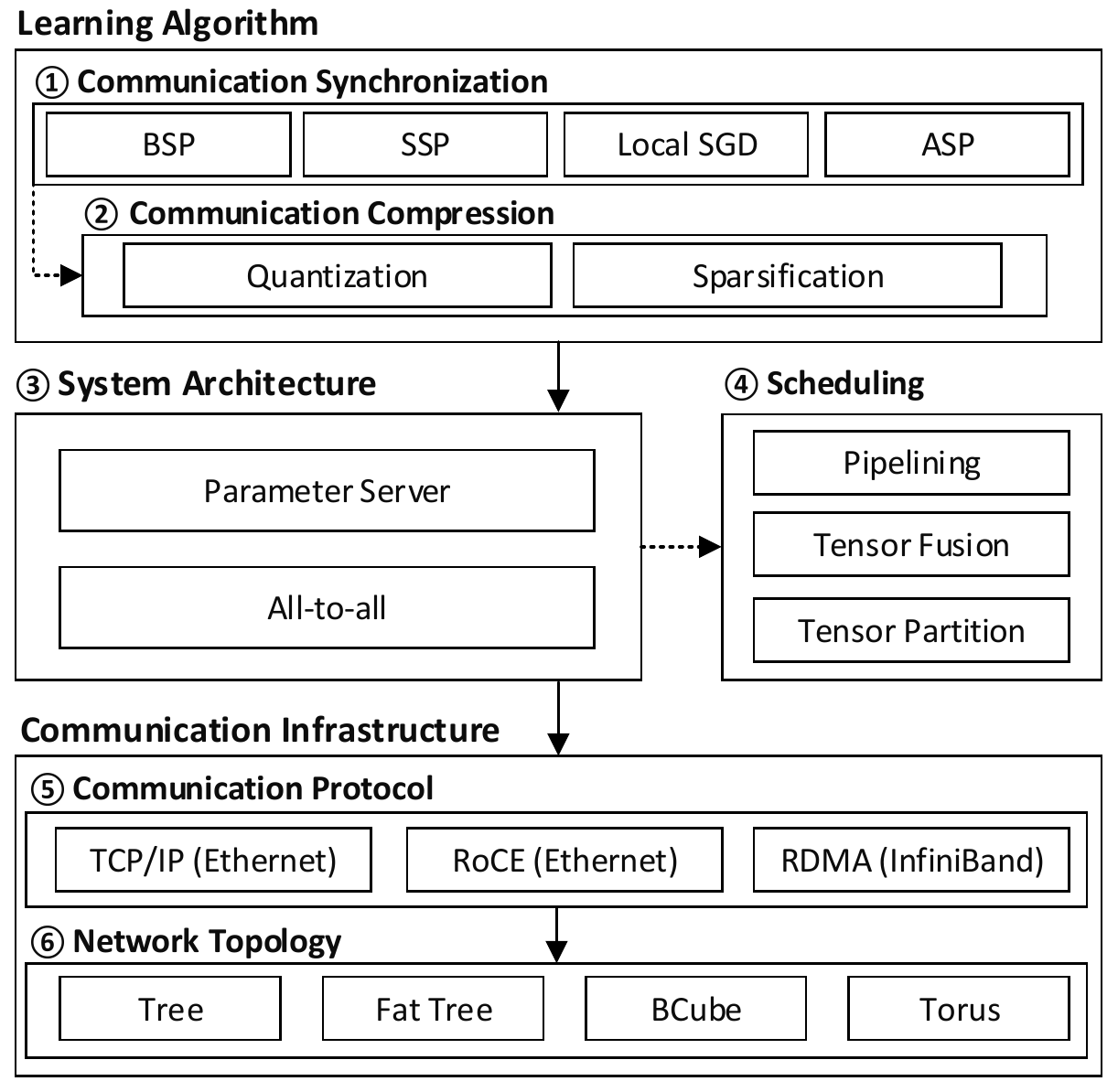}
	\caption{A three-level taxonomy of communication-efficient distributed DL.}
	\label{fig:hierarchy}
\end{figure}

\subsection{Solutions}
There have been three different directions taken to address the above challenges: 1) reducing the C2C ratio, 2) overlapping the communication tasks with the computation tasks, and 3) improving the communication performance by the advanced design of system architectures and communication primitives. In Fig. \ref{fig:hierarchy}, we develop a three-level taxonomy to describe communication-efficient distributed DL. 

\subsubsection{Learning Algorithms}
At the top, there are high-level learning algorithms with different communication complexity (aiming to reduce the C2C ratio), which can be classified into two types: 1) increasing the workload of computation (e.g., large-batch training~\cite{Lin2020Don}) and 2) reducing the communication complexity by quantization and/or sparsification. These algorithms are usually \textit{lossy} in the sense that they generate inconsistent results with the single-worker SGD. Lossy algorithms may need more iterations to achieve the same level of convergence, though each iteration completes faster.

Large-batch training is an immediate way to reduce the C2C ratio by enlarging the batch size. With proper optimization tricks (e.g., layer-wise adaptive rate scaling), large-batch training can maintain the same generalization ability as single-worker SGD. However, the local batch size is limited by the memory size of the AI processor. 

We can also relax the synchronization or reduce the communication frequency among workers (e.g., staled synchronized parallel (SSP)~\cite{zhao2019dynamic}, local SGD \cite{Lin2020Don}, and asynchronous parallel (ASP) \cite{recht2011hogwild} SGD). SSP SGD allows some workers to run more iterations before synchronization, which is efficient in a heterogeneous environment where different workers have different computing horsepower. Local SGD allows all workers to run a specific number of local updates independently before synchronization. ASP SGD enables all workers to train the model without waiting for any other workers to update the model parameters. Compression techniques such as gradient quantization \cite{bernstein2018signsgd} and sparsification \cite{shi2019distributed} are another thread of lossy algorithms. Gradient quantization quantifies each gradient into a few bits with little impact on the convergence, while gradient sparsification selects a small portion of the gradients for model updates. 

\subsubsection{System Architectures}
The middle level is the system architectures that define how the workers exchange the information. Parameter server (PS) (e.g., \cite{chen2019round}) and all-to-all (A2A) (e.g., \cite{shi2019mg}) are the two most popular system architectures, and they can be equipped with different communication scheduling algorithms that can either overlap communication with computation or improve the communication performance by tensor fusion/partition. PS is a centralized architecture that requires one or more central servers to manage the model parameters, while A2A is a decentralized architecture that exploits message passing interface (MPI) or alike to perform data communication tasks. The optimization techniques in this level are usually \textit{lossless} as they don't change the training results of the learning algorithms.

\subsubsection{Communication Infrastructure}
At the bottom level, there are diverse communication infrastructures offering the fundamental data communication services, which include communication protocols and network topologies. The optimization techniques in this level are also \textit{lossless}.

Popular communication protocols are TCP/IP, RDMA on InfiniBand, and RoCE. TCP/IP is widely supported by commodity Ethernet switches. However, it is inefficient for high-speed data communications due to the cost of data copy between the kernel buffer and the application buffer. RDMA can deliver lower latency and higher throughput than TCP/IP \cite{xue2019fast}. RDMA was originally run on InfiniBand, while RoCE (RDMA over converged Ethernet) enables the cheaper Ethernet to support RDMA. Network topology design is also important to improve the performance of distributed DL. E.g., Wang \textit{et al.}~\cite{wang2019impact} showed that BCube is more suitable than Fat-tree for distributed DL. 

In summary, a distributed training method may involve six different aspects: \circled{1} Communication Synchronization, \circled{2} Communication Compression, \circled{3} System Architecture, \circled{4} Scheduling, \circled{5} Communication Protocol, and \circled{6} Network Topology. This can be described as ``it exploits \circled{1} with/without \circled{2}, running on \circled{3} with/without \circled{4} building on \circled{5} and \circled{6}.'' In practice, BSP SGD with large-batch training is more popular than the other learning algorithms due to its good convergence property. Therefore, given a GPU cluster with a fixed communication infrastructure, the system architecture and scheduling algorithms become the key communication optimization techniques to improve the system scalability. In the next section, we continue to discuss the impact of system architectures and scheduling algorithms on the performance of distributed DL.

\section{A Popular Communication Optimization Portfolio}
\label{sec:portfolio}

\begin{figure*}[!ht]
	\centering
	\includegraphics[width=\linewidth]{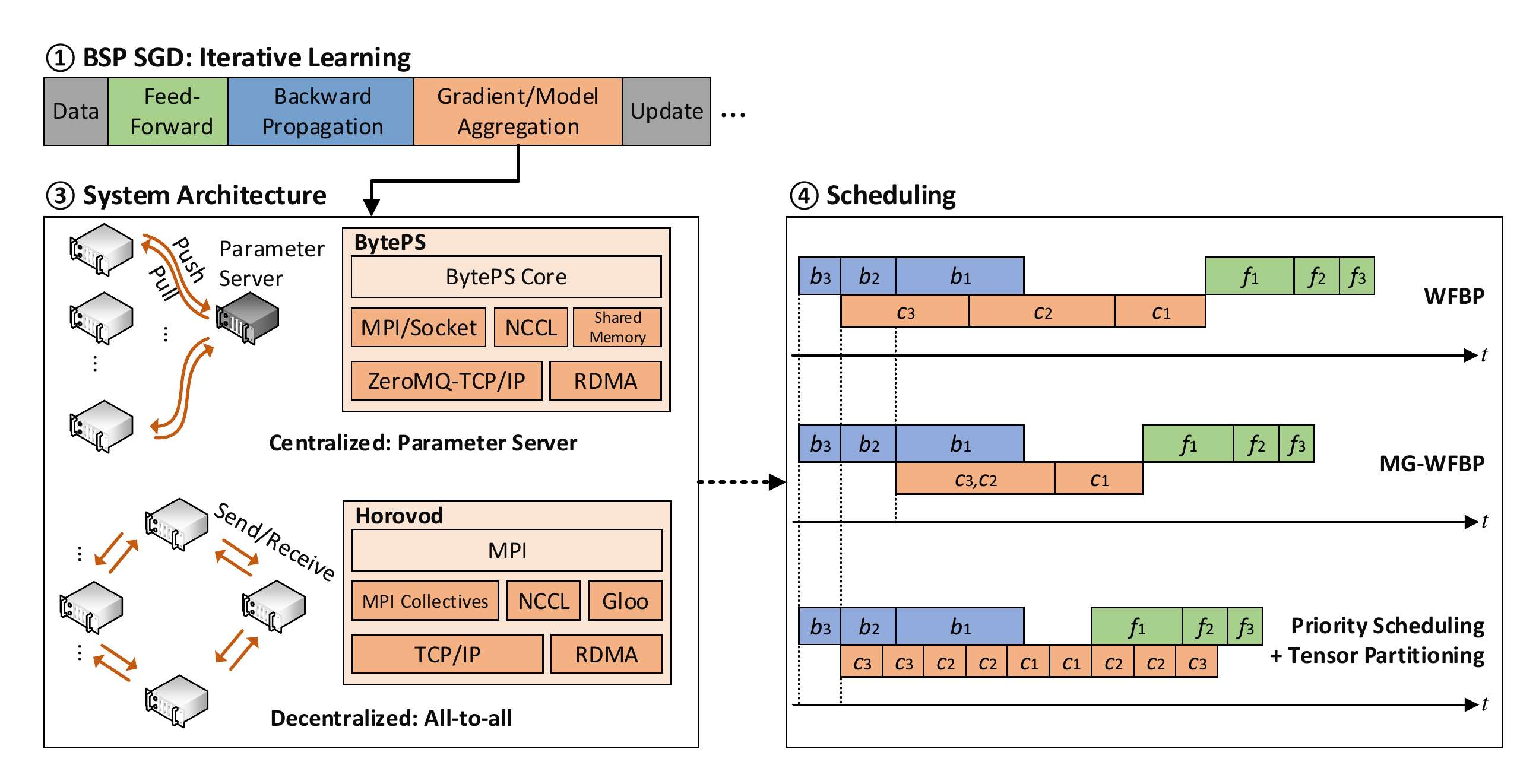}
	\caption{A communication optimization portfolio in distributed DL.}
	\label{fig:portfolio}
\end{figure*}

In this section, we focus on the communication optimization techniques in
system architecture design and scheduling algorithms. These techniques are
lossless, making them particularly appealing to industry practitioners because
model accuracy is the most important for many AI applications.
Fig.~\ref{fig:portfolio} gives an illustration of the communication
optimization techniques.

\subsection{System Architectures}
PS and A2A represent two different design philosophies, with different communication properties.

\subsubsection{Parameter Server (PS)}
In the PS architecture, a PS is logically a central server that aggregates the
gradients from the workers, updates the model parameters, and sends back the
latest model to the workers. It provides a simple and flexible
framework for the system implementation. However, since PS needs to receive
gradients from and send parameters (or averaged gradients) to all workers, it
could easily become the system bottleneck in the BSP algorithm where all
workers communicate with the PS almost simultaneously. With a single PS, the
communication traffic is $2D$ for each worker and $2ND$ for the PS. To alleviate the communication pressure on a single PS, one can deploy multiple PSes.

Here we introduce a representative PS implementation called BytePS\footnote{\url{https://github.com/bytedance/byteps}}, a highly optimized framework that supports multiple PSes by partitioning the gradient tensors in a load-balanced manner. Given $S$ PSes, the $D$-dimensional gradient is partitioned into $D/S$ parts so that each PS receives $ND/S$ gradients from $N$ workers. The received $N$ gradient tensors are averaged on the server side and sent back to all $N$ workers. Therefore, the communication traffic of each PS is reduced to $2ND/S$. 

\subsubsection{All-to-all (A2A)}
The average of the distributed gradient or parameter tensors can be calculated by an A2A operation, e.g., the all-reduce primitive in MPI. The ring-based all-reduce collective is commonly used in distributed DL, which is bandwidth optimal by dividing the tensors into small messages and exchanging those messages simultaneously in a pipelined manner. However, ring-based all-reduce has a latency term that is linear to the number of workers, which becomes inefficient for large clusters. In the high-performance communication library (NCCL\footnote{\url{https://developer.nvidia.com/nccl}}), the double binary trees algorithm \cite{sanders2009two} is integrated for dense-GPU clusters, which delivers a logarithmic latency while preserving the bandwidth optimality. For some imbalance network bandwidth systems, using the hierarchy of communication bandwidths could further improve the communication efficiency \cite{cho2019blueconnect}.

Horovod\footnote{\url{https://github.com/horovod/horovod}} is a popular distributed DL framework built for the A2A architecture and supports many state-of-the-art distributed communication libraries (e.g., MPI, NCCL, and Gloo\footnote{\url{https://github.com/facebookincubator/gloo}}).

\subsection{Scheduling}
During the training process of distributed DL, the computing and communication tasks can be described by a DAG. The layer-wise (or tensor-wise) structure of deep models makes it possible to schedule different tasks intelligently so that part of the communication cost can be hidden, as shown in Fig.~\ref{fig:portfolio} \circled{4}.

\subsubsection{Layer-wise Pipelining and Tensor Fusion} 
A deep model consists of a stack of layers, and the learnable parameters of
each layer are generally represented by one or two tensors. During the
backpropagation, if the gradients of layer $P$ have been computed, then they can
be immediately communicated so that the communication task can be pipelined with the computing task of layer
$P-1$. The naive pipelining between communications and computations during
backpropagation is also called wait-free backpropagation (WFBP)
\cite{zhang2017poseidon}, which can be applied to both PS and A2A
architectures. 

In A2A with pipelining, an all-reduce operation is required for each tensor, which usually divides the tensor into multiple small messages. Considering that transmitting two small messages together is generally faster than transmitting the two messages separately (e.g., in our 100Gbps InfiniBand cluster, transmitting a 16KiB message takes 7.85us, while transmitting a 32KiB message takes 10.1us), the MG-WFBP algorithm adopts the idea of tensor fusion by optimally merging the gradients of several consecutive layers to minimize the iteration time~\cite{shi2019mg}. Tensor fusion can effectively alleviate the negative impact of transmitting small messages.

\subsubsection{Tensor Partitioning and Priority Scheduling}
In the PS architecture, the communication happens between a worker and a PS
and a tensor can be transmitted as a single message, making tensor fusion less
beneficial than in A2A. Other than pipelining, there is another opportunity
for performance improvement by priority scheduling. In PS, there are two
directions of communications: push of gradients and pull of parameters.
For each layer, the pull of parameters is commonly followed
by the push of gradients. If the current layer has a large tensor, it
would block other layers with small tensors. ByteScheduler~\cite{peng2019generic} 
is the efficient scheduling strategy that partitions a large tensor into multiple
smaller ones and allows the lower layers to be scheduled ahead of the higher 
layers. By using the priority scheduling, it is possible to overlap the communication
tasks with both feed-forward and backpropagation computing tasks~\cite{jayarajan2019priority,peng2019generic}.

\section{Comparative Study}
\label{sec:experiments}

To demonstrate the key factors that affect the scalability of the optimization
portfolio presented in Section~\ref{sec:portfolio}, we evaluate and compare
the system performance of seven representative distributed training methods
listed in Table~\ref{table:comparedmethods}, which are widely used in practice
and serve as good examples to quantitatively study different optimization
techniques. BSP-PS and BSP-A2A are the baseline cases without special
optimization, which are used to compare the efficiency of PS and A2A. WFBP-PS and
WFBP-A2A are with WFBP scheduling, which can evaluate the effectiveness of
WFBP on different architectures. MG-WFBP uses tensor fusion to address the
latency problem of WFBP-A2A. ByteScheduler-PS and ByteScheduler-A2A are with
both pipelining and tensor partition under PS and A2A architectures
respectively, which can show the performance of tensor partition.

We choose three representative deep models for evaluation, namely ResNet-50,
BERT-Base, and BERT-Large, which are commonly used in image classification and
natural language processing. Their model intensities are 470, 249, and 248,
respectively. On RTX2080Ti, ResNet-50 and BERT-Base can support a local batch
size of 64, while BERT-Large can only support 8. These three
models can well illustrate the impact of model intensity and batch size on the
system scalability. 

\begin{table*}[!ht]
	\centering
	\caption{Experimental Settings for Evaluation. For BytePS, as suggested by the official release, we use multiple PSes whose amount is the same as the number of worker servers. Each worker server has multiple workers (i.e., GPUs).}
	\label{table:comparedmethods}
	\begin{tabular}{|c|c|c|c|c|c|c|}
	\hline
\multirow{2}{*}{Method} & System Architecture &  \multicolumn{3}{c|}{Scheduling}  & \multirow{2}{*}{\shortstack[c]{Distributed\\Software}} &\multirow{2}{*}{\shortstack[c]{Common\\Libraries}}\\
	 & PS/All-to-all & Pipelining & Tensor Fusion & Tensor Partition &  &  \\\hline\hline
	BSP-PS \cite{zhang2017poseidon}  & PS & \xmark  & \xmark & \xmark  & BytePS &\multirow{7}{*}{\shortstack[c]{PyTorch-1.4\\CUDA-10.1\\NCCL-2.4.8}} \\\cline{1-6}
	BSP-A2A \cite{shi2019mg,sanders2009two} & All-to-all  & \xmark & \xmark & \xmark & Horovod &\\\cline{1-6}
	WFBP-PS \cite{zhang2017poseidon} & PS & \cmark  & \xmark & \xmark  & BytePS&  \\\cline{1-6}
	WFBP-A2A \cite{shi2019mg,sanders2009two} & All-to-all  & \cmark & \xmark & \xmark &  Horovod &\\\cline{1-6}
	MG-WFBP \cite{shi2019mg} & All-to-all & \cmark & \cmark & \xmark & Horovod &\\\cline{1-6}
	ByteScheduler-PS \cite{peng2019generic} & PS & \cmark & \xmark &\cmark  &  BytePS &\\\cline{1-6}
	ByteScheduler-A2A \cite{peng2019generic} & All-to-all & \cmark & \xmark &\cmark  &  Horovod &\\\hline
	\end{tabular}
\end{table*}

\subsection{Experimental Settings}
\noindent \textbf{Hardware: } We conduct experiments on a GPU cluster with
RDMA over 100Gbps IB. The cluster consists of 8 nodes (or worker servers). Each node has four Nvidia RTX2080Ti GPUs (11GB RAM) interconnected by PCIe3.0 x16, two Intel(R) Xeon(R) Gold 6230 CPUs, and 512GB memory. 

\vspace{.4em}
\noindent \textbf{Software: }We exploit PyTorch-1.4\footnote{\url{https://pytorch.org/}} as the backbone framework with GPU libraries of CUDA-10.1, cuDNN-7.6 and NCCL-2.4.8. We use the highly optimized libraries of BytePS and Horovod for PS and A2A architectures, respectively. 

\vspace{.4em}
\noindent \textbf{Measurements: }We use the metric of system throughput (i.e., samples per second) in processing the data samples to evaluate the performance. For ResNet-50, a sample is an image with a resolution of $224\times224\times3$; for BERT-Base and BERT-Large, a sample is a sentence with a length of $64$ words. We use the SGD training with a single RTX2080Ti as the baseline to calculate the speedup. Note that when comparing the results between different number of workers, they have different effective batch sizes and their convergence might be different.

\subsection{Experimental Results}
\begin{figure*}[!ht]
	\centering
	\subfigure[ResNet-50 ($I$ = 470, $LBS$ = 64)]
	{
    	\includegraphics[width=0.32\linewidth]{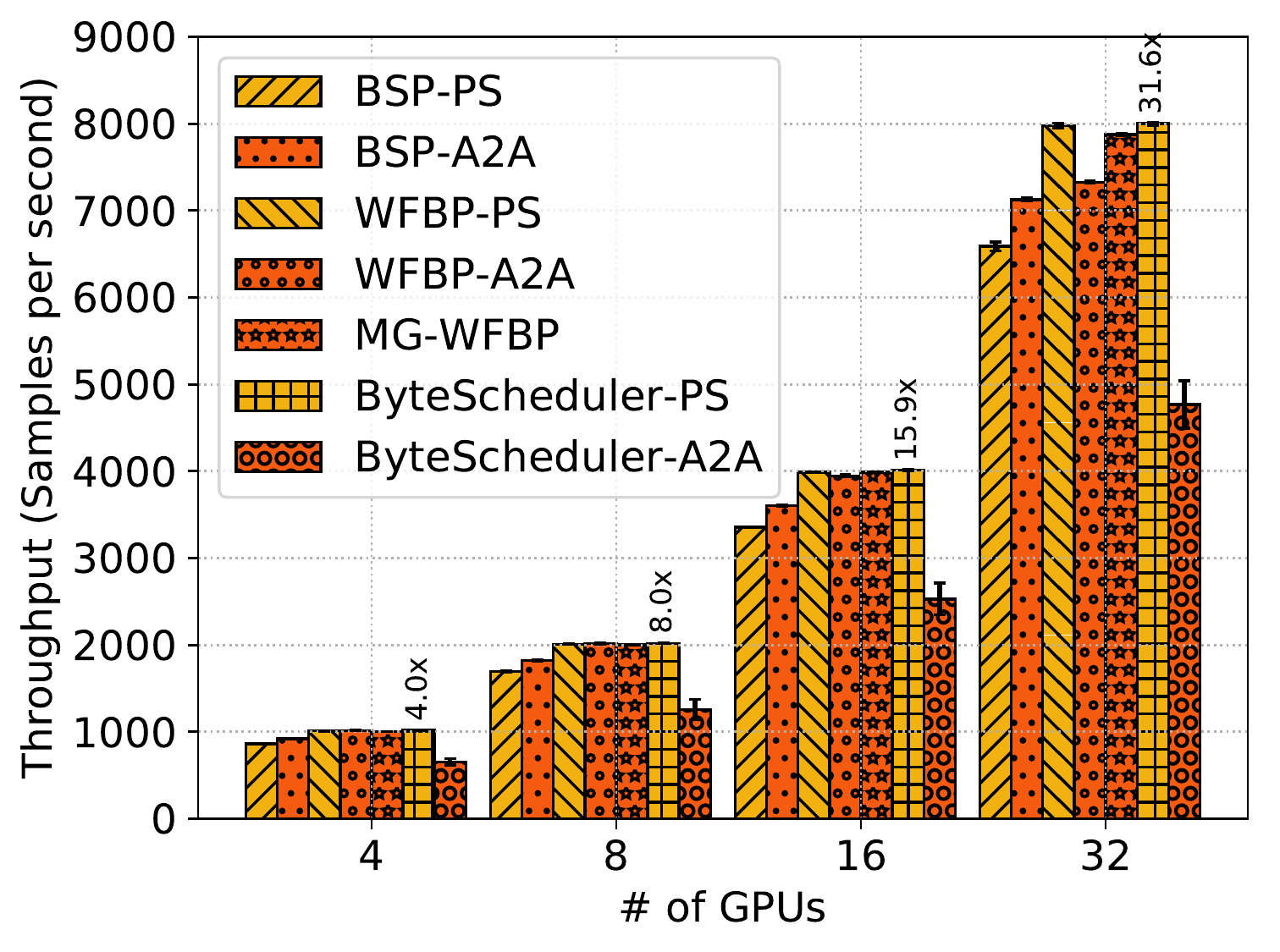}
	}\hspace{-2pt}
	\subfigure[BERT-Base ($I$ = 249, $LBS$ = 64)]
	{
    	\includegraphics[width=0.32\linewidth]{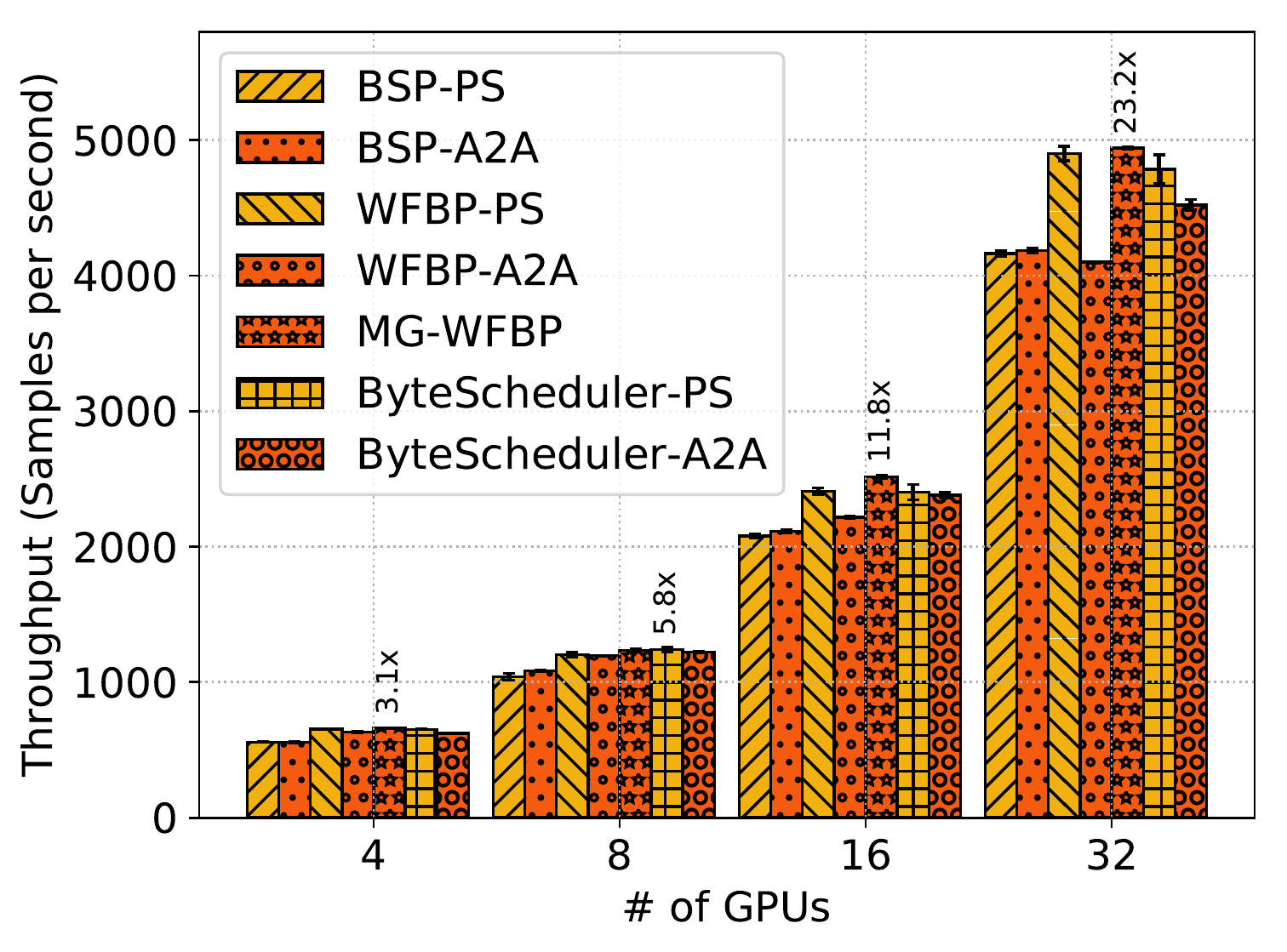}
	}\hspace{-2pt}
	\subfigure[BERT-Large ($I$ = 248, $LBS$ = 8)]
	{
    	\includegraphics[width=0.32\linewidth]{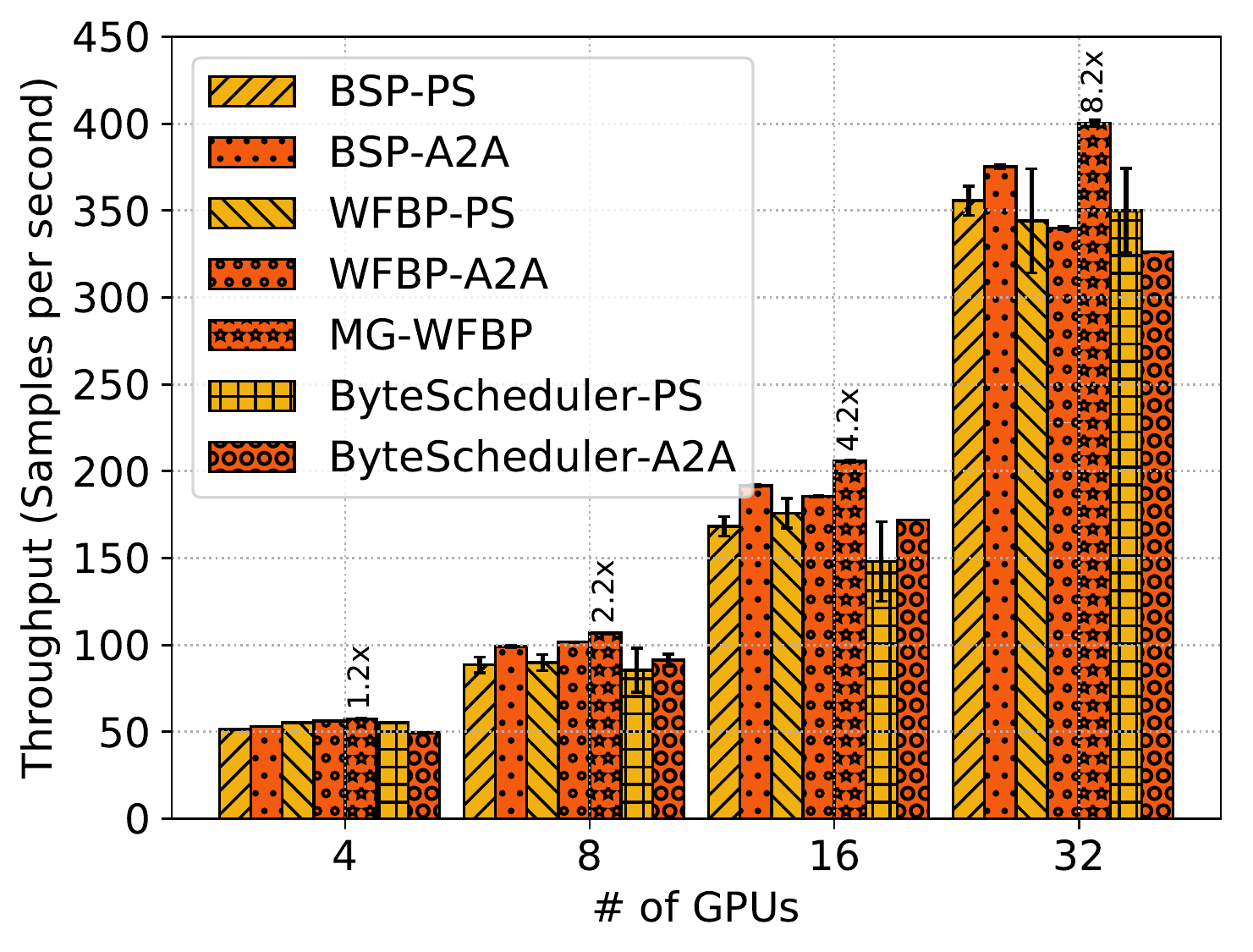}
	}
	\caption{System throughput comparison. $I$: model intensity. $LBS$: local batch size. The numbers on the top of the bars are the best speedups among the seven evaluated methods over the single-GPU SGD algorithm.}
	\label{fig:results}
\end{figure*}

Fig.~\ref{fig:results} depicts the experimental results, averaged over five independent experiments. For each run, we conduct 10 training iterations for warm-up, and another 100 iterations for measuring the average throughput. We summarize our major findings in Table~\ref{table:findings}.

\begin{table*}[!ht]
\centering
\caption{Major Findings of Experimental Results.}\label{table:findings}
    \begin{tabular}{|c|l|l|}
    \hline
    Section & Related Factors & Major Findings \\\hline\hline
    \ref{subsubsec:exp:intensity} & Model intensity and  & 1) The model with higher model intensity is easier to be parallelized.\\
    & batch size & 2) Increasing the batch size to reduce the C2C ratio makes the parallelism easier, but the maximum local\\
    &  &  batch size is limited by GPU memory. \\\hline
    \ref{subsubsec:exp:psvsa2a} & PS vs. A2A & 3) There is no single winner in PS and A2A. Both can achieve comparable performance when enhanced \\
    &  & with different optimization algorithms. But PS needs extra servers and network switch ports to be \\
    &  & competitive with A2A.\\\hline
    \ref{subsubsec:exp:scheduling} & Scheduling & 4) Wait-free backpropagation (WFBP) can generally hide some communication costs. Scheduling is \\
    &  & helpful when the communication time is comparable to the computing time per worker.\\
    &  & 5) Tensor fusion (e.g., MG-WFBP) is suitable for A2A because it addresses the inefficiency of transmitting \\
    &  & small messages in A2A.\\
    &  & 6) Tensor partition (e.g., ByteScheduler) is suitable for PS, which makes communications better \\
    &  & overlapped with computations.\\\hline
    \end{tabular}
\end{table*}

\subsubsection{Impact of Model Intensity and Batch Size}\label{subsubsec:exp:intensity}
\noindent \textbf{ResNet-50 vs. BERT-Base: } As the model intensity of ResNet-50 is about twice as large as BERT-Base, and their local batch sizes are both 64, the C2C ratio of ResNet-50 is around half of BERT-Base. Comparing Fig.~\ref{fig:results}(a) with Fig.~\ref{fig:results}(b), we see that ResNet-50 has much better scalability than BERT-Base. For example, on the intra-node training with 4 GPUs, we can achieve an optimal speedup of $4\times$ on ResNet-50, but only $3.1\times$ on BERT-Base; with 32 GPUs, ResNet-50 has a speedup of $31.6\times$, while BERT-Base has only $23.2\times$. The results confirm that a model with higher intensity is easier to be parallelized.

\vspace{.4em}
\noindent \textbf{BERT-Base vs. BERT-Large: } The model intensities of BERT-Base and BERT-Large are very close, but the local batch size for BERT-Base is $8\times$ larger than BERT-Large due to the smaller GPU memory footprint. Therefore, the C2C ratio of BERT-Large is about $8\times$ higher than BERT-Base, which makes BERT-Large much more difficult to be parallelized, as confirmed by comparing Fig.~\ref{fig:results}(c) with Fig.~\ref{fig:results}(b). The smaller speedups of BERT-Large are mainly due to the small batch size and limited bandwidth of PCIe3.0. For example, 4-GPU training on BERT-Large has a maximum of $1.2\times$ speedup, while it is $3.1\times$ for BERT-Base. The small GPU memory size of RTX2080Ti and the limited bandwidth of PCIe3.0 are not suitable for distributed training of BERT-Large. For comparison, when training BERT-Large on a much more expensive server with four Nvidia V100 GPUs (with 32GB memory) interconnected by NVLink (with more than $10\times$ higher bandwidth than PCIe3.0), the local batch size can be as large as 128, and we achieved a speedup of $3.82\times$. 

\subsubsection{System Architecture: PS vs. A2A}\label{subsubsec:exp:psvsa2a}
It is well known that the PS architecture with a single PS does not scale well. In our evaluation on the PS architecture, we use the same number of PSes and worker servers \cite{peng2019generic}. Notice that, in this setting, the PS architecture consumes more network switch ports and more total network bandwidth than A2A. 

Regarding BSP-PS and BSP-A2A without pipelining, BSP-A2A outperforms BSP-PS in all cases. However, when exploiting WFBP~\cite{zhang2017poseidon} to pipeline communications with computations, WFBP-PS outperforms WFBP-A2A, especially on 32 workers. This is because the A2A architecture has a non-negligible latency term that is logarithmic/linear to the number of workers with tree/ring-based algorithms, and WFBP requires the gradients aggregated tensor-wisely, resulting in noticeable startup overheads~\cite{shi2019mg}. The tensor fusion technique~\cite{shi2019mg} can well address this startup problem. As we observe from Fig.~\ref{fig:results}, MG-WFBP achieves the best speedup on BERT-Base (except the case of 8 workers) and BERT-Large. But for ResNet-50 with higher model intensity, ByteScheduler-PS performs slightly better than MG-WFBP. In summary, there is no clear winner between PS and A2A. Both architectures can achieve comparable performance when equipped with suitable optimization techniques. However, PS needs extra servers and switch ports to keep the competitive edge with A2A.

\subsubsection{Scheduling}\label{subsubsec:exp:scheduling}
The idea of scheduling is to overlap communication tasks with computing tasks.
Regarding the WFBP algorithm, in most cases WFBP-PS and WFBP-A2A both run
faster than BSP-PS and BSP-A2A, respectively. But WFBP-A2A sometimes suffers
from the startup latency problem as many small messages need to be transferred,
e.g., under the case of BERT-Base and BERT-Large with 32 workers. MG-WFBP
significantly improves the scalability of WFBP-A2A, especially with a large
number of workers. ByteScheduler-A2A schedules the communications in the
opposite direction with MG-WFBP by partitioning tensors instead of merging
tensors, and its performance is not very promising. However, with the PS
architecture, ByteScheduler-PS slightly outperforms WFBP-PS in ResNet-50.
This indicates that without bringing extra heavy latency by partitioning
tensors, communications of partitioned tensors can be better scheduled to
overlap with backpropagation and feed-forward computations
\cite{peng2019generic}. In summary, scheduling algorithms can 
improve the system scalability by hiding the communication overhead. However, when the
communication time dominates the training time (e.g., BERT-Large), the overall
speedup becomes rather limited and we need to either improve the network
speed or consider lossy algorithms.

\section{Challenges and Future Directions}
Despite many techniques are proposed to address the communication problem in distributed DL, some technical challenges remain open to answer.

\subsection{Communication Compression}
As the model size increases, the communication cost grows,
which could result in a very high C2C ratio. Lossless optimization algorithms in system
architecture design and scheduling can only achieve marginal improvement since
the communication cost dominates the training time. The communication
compression techniques would be useful to significantly
reduce the communication traffic in such cases. The primary challenge is 
how to maintain the model accuracy while keeping the
convergence performance. Existing methods have proven that communication
compression can achieve the same asymptotic convergence speed as vanilla SGD.
Yet in practice, with a very high compression ratio, it generally requires
more iterations to achieve the target loss value. One possible direction is to
set different compression ratios for different layers to maximize the
exchanged information. Another possibility is to dynamically set
appropriate compression ratios at different training iterations.  

\subsection{Automatically Selected System Architecture}
The PS and A2A architectures are widely deployed for the BSP algorithm in both
industry and academia. Intuitively, the A2A architecture is more efficient
than PS as it requires no central servers; but A2A is more latency-sensitive than PS. Furthermore, one can use multiple PSes to reduce the
central server's network footprint. More uncertainly, with different hardware
configurations, model properties, and scheduling algorithms, no solution is always
better in all cases. An interesting yet challenging problem is to build
mathematical performance models for both PS and A2A according to the training
environments (e.g., the number of GPUs, network topology, link bandwidth and
latency, model properties, etc.), so that a better architecture can be chosen
for training the target model.

\subsection{Generic Scheduling}
According to the characteristics of distributed DL, various scheduling
algorithms try to maximize the parallelism of computing tasks
and communication tasks. However, these algorithms were built upon the DAG of
BSP with three types of tasks (i.e., feed-forward,
backpropagation, and gradient communication). The
scheduling algorithm only brings marginal improvement if the communication
time is much longer than the computing time. Although communication
compression can reduce the communication cost, current scheduling methods are not directly applicable to the
BSP with gradient compression because compression
introduces extra non-negligible computational costs and smaller communication
traffic, which makes the scheduling more difficult. One possible solution is
to design a generic scheduler for configured DAGs. The DAG would be
changed due to tensor partition or fusion. For the configured DAG, the
scheduler can use some heuristic algorithms to search for the configuration
with better performance. Furthermore, current scheduling techniques such as
MG-WFBP~\cite{shi2019mg} and ByteScheduler~\cite{peng2019generic} take two
opposite directions (i.e., tensor fusion and tensor partition) for scheduling.
In practice, no one is always better. An intelligent scheduler should be adaptive
to the training environment and dynamically determine whether the tensors
should be merged or partitioned to achieve higher performance.

\section{Conclusion}
In this article, we gave an overview of the techniques to address the communication challenges in distributed deep learning. We first analyzed the communication problems in distributed training of deep learning models, and then presented a taxonomy and survey of the existing state-of-the-art technologies. We particularly focused on the commonly used lossless methods and provided a quantitative analysis to these methods based on real-world experiments. Finally, we discussed the challenges and possible future research directions in this area.

\section*{Acknowledgments}
The research was supported in part by Hong Kong RGC GRF grants under the contracts HKBU 12200418, HKUST 16206417 and 16207818, and in part by National Natural Science Foundation of China under Grant 62002240.

\bibliography{cites}
\bibliographystyle{IEEEtran}

\section*{Biographies}
\renewenvironment{IEEEbiography}[1]
  {\IEEEbiographynophoto{#1}}
  {\endIEEEbiographynophoto}
  
\vspace{-20pt}
\begin{IEEEbiography}{Shaohuai Shi} (shaohuais@cse.ust.hk) received a B.E. degree in software engineering from South China University of Technology, P.R. China, in 2010, an MS degree in computer science from Harbin Institute of Technology, P.R. China in 2013, and a Ph.D. degree in computer science from Hong Kong Baptist University in 2020. He is currently a research assistant professor in the Department of Computer Science and Engineering at the Hong Kong University of Science and Technology. His research interests include GPU computing and machine learning systems. He is a member of the IEEE.
\end{IEEEbiography}
\vspace{-20pt}
\begin{IEEEbiography}{Zhenheng Tang} (zhtang@comp.hkbu.edu.hk) received a B.E. degree in communication engineering from Huazhong University Of Science and Technology, P.R. China, in 2018. He is a Ph.D. student at Hong Kong Baptist University. His research interests include GPU computing and distributed deep learning. 
\end{IEEEbiography}
\vspace{-20pt}
\begin{IEEEbiography}{Xiaowen Chu} (chxw@comp.hkbu.edu.hk) received a B.E. degree in computer science from Tsinghua University, P.R. China, in 1999, and a Ph.D. degree in computer science from The Hong Kong University of Science and Technology in 2003. Currently, he is a full professor in the Department of Computer Science, Hong Kong Baptist University. His research interests include parallel and distributed computing, cloud computing and wireless networks. He is serving as an Associate Editor of IEEE Access and IEEE Internet of Things Journal. He is a senior member of the IEEE.
\end{IEEEbiography}
\vspace{-20pt}
\begin{IEEEbiography}{Chengjian Liu} (liuchengjian@sztu.edu.cn) received his MS degree in College of Computer Science and Software Engineering, Shenzhen University, P.R. China, in 2013, and his Ph.D. degree in computer science from the Hong Kong Baptist University in 2018. Currently, he is an assistant professor in the College of Big Data and Internet, Shenzhen Technology University. His research interests include Distributed Storage, Blockchain, General-Purpose GPU Computing.
\end{IEEEbiography}
\vspace{-20pt}
\begin{IEEEbiography}{Wei Wang} (weiwa@cse.ust.hk) received his B.Eng. (Hons.) and M.Eng. degrees from Shanghai Jiao Tong University, and a Ph.D. degree from the University of Toronto in 2015, all in the Department of Electrical and Computer Engineering. He is an Assistant Professor in the Department of Computer Science and Engineering at the Hong Kong University of Science and Technology (HKUST). He is also affiliated with HKUST Big Data Institute. His research interests cover the broad area of distributed systems, with special emphasis on big data and machine learning systems, cloud computing, and computer networks in general.
\end{IEEEbiography}
\vspace{-20pt}
\begin{IEEEbiography}{Bo Li} (bli@cse.ust.hk) received a BEng in computer science from Tsinghua University, Beijing and a Ph.D. degree in electrical and computer engineering from the University of Massachusetts at Amherst. He is a professor in the Department of Computer Science and Engineering, Hong Kong University of Science and Technology. He was the chief technical advisor of ChinaCache Corp. (NASDAQ CCIH), the largest CDN operator in China. He was a Cheung Kong visiting chair professor with Shanghai Jiao Tong University (2010-2013) and an adjunct researcher with Microsoft Research Asia (1999-2007) and with Microsoft Advance Technology Center (2007-2009). His current research interests include: multimedia communications, the Internet content distribution, datacenter networking, cloud computing, and wireless sensor networks. He made pioneering contributions in the Internet video broadcast with the system, Coolstreaming, which was credited as the world first large-scale Peer-to-Peer live video streaming system. The work appeared in IEEE INFOCOM (2005) received the IEEE INFOCOM 2015 Test-of-Time Award. He has been an Editor or a Guest Editor of more than a dozen of the IEEE journals and magazines. He was the co-TPC chair of the IEEE INFOCOM 2004. He received five Best Paper Awards from the IEEE. He received the Young Investigator Award from Natural Science Foundation of China (NFSC) in 2005, the State Natural Science Award (2nd Class) from China in 2011. He is a fellow of the IEEE.
\end{IEEEbiography}

%
%
%
%

\end{document}